\documentclass[preprint,showpacs,preprintnumbers,amsmath,amssymb]{revtex4}

\usepackage{graphicx}
\usepackage{dcolumn}
\usepackage{bm, amsmath,amssymb, mathrsfs, comment, color, mathtools}
\newcommand{\na}{\nabla}
\newcommand{\lt}{\left}
\newcommand{\rt}{\right}
\newcommand{\cb}{\underline{c}}

\newcommand{\rw}{\rightarrow}

\begin{document}

\preprint{}

\title{Supertranslation invariance of angular momentum}
\author{Po-Ning Chen}
 \affiliation{%
University of California Riverside, Department of Mathematics, 900 University Avenue, Riverside, California 92521, USA\\
}%

\author{Mu-Tao Wang}
 \affiliation{%
Columbia University, Department of Mathematics, 2990 Broadway, New York, New York 10027, USA\\
}%
\author{Ye-Kai Wang}
 \affiliation{%
National Cheng Kung University, Department of Mathematics, 1 Dasyue Road, Tainan City 70101, Taiwan\\
}%

\author{Shing-Tung Yau}%
\affiliation{%
Harvard University, Department of Mathematics, One Oxford Street, Cambridge, Massachusetts 02138, USA\\
}%

\date{\today}

\begin{abstract}

LIGO's successful detection of gravitational waves has revitalized the theoretical understanding of the angular momentum
carried away by gravitational radiation. An infinite dimensional supertranslation ambiguity has presented an essential difficulty for decades of study. Recent advances were made to address and quantify the supertranslation ambiguity in the context of compact binary coalescence. Here we present the first definition of angular momentum in general relativity that is completely free from supertranslation ambiguity. The new definition was derived  from the limit of the quasilocal angular momentum defined previously by the authors. A new definition of center of mass at null infinity is also proposed and shown to be supertranslation invariant. Together with the classical Bondi-Sachs energy-momentum, they form a complete set of conserved quantities at null infinity that transform according to basic physical laws.
\end{abstract}

\pacs{}
\maketitle

\section{\label{sec:level1} Introduction}
The definitions of conserved quantities such as mass and angular momentum have been among the most difficult problems since the genesis of general relativity. According to Einstein's equivalence principle, there is no density for gravitation and no canonical coordinate system for spacetime. The issue is further complicated by the nonlinear nature of Einstein's eponymous equation. One of the most important problems is the definition of angular momentum for a distant observer, or angular momentum at null infinity, a notion that has been studied for
decades \cite{AM, AS, Rizzi}.  An essential difficulty is presented by the ambiguity of supertranslations, an infinite dimensional subgroup of  the Bondi-Metzner-Sachs (BMS) group \cite{BS}. 
According to Penrose \cite{Penrose}, the very concept of angular momentum gets shifted by these supertranslations and ``it is hard to see in these circumstances how one can rigorously discuss such questions as the angular momentum
carried away by gravitational radiation" (page 654 of \cite{Penrose}).  The astronomical event GW150914 \cite{GW} observed by LIGO corresponds to the coalescence of a binary black hole, and recent advances  \cite{ADK} were made to address and quantify the supertranslation ambiguity in general compact binary coalescences. Nevertheless, to deal with general isolated systems it is desirable to have a rigorous definition that is free from any supertranslation ambiguity. In this article, we propose the first definition of angular momentum at null infinity that is supertranslation invariant. The definition is derived as the limit of quasilocal angular momentum that was proposed in \cite{CWY2}  and evaluated at null infinity in \cite{KWY-CWY1}. Comparing with existing definitions, the new definition contains an important correction term (that has never appeared in any previous definitions), which comes from solving the optimal isometric embedding equation in the theory of Wang-Yau quasilocal mass \cite{WY}. This provides the reference term  that is critical in the Hamiltonian approach of defining conserved quantities. Our theory also produces a definition of center of mass at null infinity which is shown to be supertranslation invariant as well.

\section{\label{sec:level1} Conserved quantities in Bondi-Sachs coordinates}

We consider a  Bondi-Sachs coordinate system $(u, r,  x^2, x^3)$ in which the physical spacetime metric takes the form
\begin{equation}\label{spacetime_metric} -UV du^2-2U dudr+r^2 h_{AB}(dx^A+W^A du)(dx^B+W^B du), A, B=2, 3.\end{equation} The future null infinity $\mathscr{I}^+$ corresponds to the idealized null hypersurface $r=\infty$ and can be viewed $\mathscr{I}^+=I\times S^2$ with coordinates $(u, x)$ where $u\in I$ and $x=(x^2, x^3)\in S^2$, the unit sphere. The outgoing radiation condition \cite{Sachs} implies the following expansions in inverse powers of $r$:
 \[\begin{split}
V&=1-\frac{2m}{r}+  O(r^{-2}),\\
W^A&= \frac{1}{2r^2} \na_B C^{AB} + \frac{1}{r^3} \lt( \frac{2}{3}N^A - \frac{1}{16} \na^A |C|^2 -\frac{1}{2} C^{AB} \na^D C_{BD} \rt) + O(r^{-4}),\\
h_{AB}&={\sigma}_{AB}+\frac{C_{AB}}{r}+ O(r^{-2}),\end{split} \] where $\sigma_{AB}(x)$ is a standard round metric on $S^2$ and $\nabla_A$ denotes the covariant derivative with respect to $\sigma_{AB}$. The indices are contracted, raised, and lowered with respect to the metric $\sigma_{AB}$.
Defining on $\mathscr{I}^+$ $(r=\infty)$ are the {\it mass aspect} $m=m(u, x)$, the {\it angular aspect} $N_A = N_A(u, x)$, and the {\it shear} $C_{AB}=C_{AB}(u, x)$ of this Bondi-Sachs coordinate system.  We also define the {\it news} $N_{AB} = \partial_u C_{AB}$.  

The standard formulae for the Bondi-Sachs energy-momentum \cite{BS} at a $u$ cut along $\mathscr{I}^+$ are 
\begin{equation}\label{EM} E (u) = \int_{S^2} 2m (u, \cdot), P^k (u) = \int_{S^2} 2m (u, \cdot) \tilde{X}^k, k=1, 2, 3\end{equation}
where $\tilde{X}^k, k=1, 2 ,3$ are the standard coordinate functions on $\mathbb{R}^3$ restricted to $S^2$.

In order to define the angular momentum, we consider the decomposition of $C_{AB}$ into 
\begin{equation}\label{CAB}C_{AB}=\nabla_A\nabla_B c-\frac{1}{2} \sigma_{AB} \Delta c+\frac{1}{2}(\epsilon_A^{\,\,\,\, E} \nabla_E \nabla_B \underline{c}+\epsilon_B^{\,\,\,\, E} \nabla_E \nabla_A \underline{c})\end{equation} where $\epsilon_{AB}$ denotes  the volume form of $\sigma_{AB}$.  $c=c(u, x)$ and $\underline{c}=\underline{c}(u, x) $ are the closed and co-closed potentials of $C_{AB}(u, x)$. They are chosen to be of $\ell\geq 2$ harmonic modes and thus such a decomposition is unique. 

The asymptotic symmetry of $\mathscr{I}^+$ consists of the BMS fields \cite{Sachs, ADK}.  We say a BMS field $Y$ is a {\it rotation BMS field} if in a Bondi-Sachs coordinate system
$(u, x)$, 
\begin{equation}\label{rotation_BMS}Y=\hat{Y}^A\frac{\partial}{\partial x^A}\end{equation} where $\hat{Y}^A(x)$ is a rotation Killing field on $S^2$. We define the angular momentum with respect to a rotation BMS field $Y$ in the following:

\noindent \textit{Definition of angular momentum:}
For a rotation BMS field $Y$ that is tangent to $u$ cuts on $\mathscr{I}^+$, the  angular momentum of a $u$ cut is defined to be :
\begin{align}\label{AM}
 J (u, Y)= \int_{S^2}  Y^A \lt( N_{A} -\frac{1}{4}C_{AB}\nabla_{D}C^{DB} - c\nabla_{A}m   \rt) (u, \cdot)
\end{align}

Suppose  $(\bar{u}, \bar{x})$ is another Bondi-Sachs coordinate system, we define similarly: 
\begin{equation}
{J} (\bar{u}, \bar{Y})=\int_{S^2}    \bar{Y}^A \left(\bar{N}_A-\frac{1}{4}\bar{C}_{A}^{\,\,\,\,D}\bar{\nabla}^B \bar{C}_{DB}     - \bar{c}\bar{\nabla}_{A}
\bar{m}         \right) (\bar u, \cdot),
\end{equation} where $\bar{Y}$ is a rotation BMS field that is tangent to $\bar{u}$ cuts.

A {\it supertranslation} is a change of Bondi-Sachs coordinates $(\bar{u}, \bar{x})\rightarrow (u, x)$ such that 
\begin{equation}\label{coord_change} u = \bar u + f (x), x=\bar{x}\end{equation} on $\mathscr{I}^+$ for a function $f$ that is defined on $S^2$. Two rotation BMS fields $\bar{Y}$ and ${Y}$ are said to be related by the supertranslation $f$ if, in the $(u, x)$ coordinate system\begin{equation}\label{BMS_related}\bar{Y}=\hat{Y}^A\frac{\partial}{\partial x^A}+\hat{Y}(f)\frac{\partial}{\partial u}
\text{ and }  Y=\hat{Y}^A \frac{\partial}{\partial x^A}\end{equation} for a rotation Killing field $\hat{Y}$ on $S^2$. In this case, $\bar{Y}$ is tangent to the $\bar{u}$ cuts while ${Y}$ is tangent to the ${u}$ cuts.

We show (Theorem 1) that the total fluxes of $J (u, Y)$ and ${J} (\bar{u}, \bar{Y})$ are the same when $Y$ and $\bar{Y}$ are related by a supertranslation $f$ of harmonic mode $\ell\geq 2$, thus removing the supertranslation ambiguity.

The expression \eqref{AM} originated from the definition of quasilocal angular momentum, which was proposed in \cite{CWY2} to complement the definitions of quasilocal mass-energy-momentum in \cite{WY}. Their limits at null infinity were evaluated in \cite{KWY-CWY1}. The expression $ \int_{S^2}  Y^A \lt( N_{A} -\frac{1}{4}C_{AB}\nabla_{D}C^{DB}\rt)(u, \cdot)\coloneqq \tilde{J}(u, Y)$ in \eqref{AM} already appeared in other definitions of angular momentum \cite{AM, CJK}, while the last term in \eqref{AM} that involves $c$ is new and plays an indispensable role in supertranslation invariance. This term arises  naturally from solving the optimal isometric embedding in the theory of Wang-Yau quasilocal mass \cite{WY} and provides the reference term that is critical in the Hamiltonian approach of defining conserved quantities. Rizzi's angular momentum definition \cite{Rizzi} is essentially $\tilde{J}(u, Y)$ in the framework of \cite{Ch}. Without a suitable reference term, his definition is only valid for a restricted class of foliations at null infinity, and does not satisfy the supertranslation invariance property.

\section{Supertranslation invariance of total fluxes}

\begin{figure}[h]
\caption{A supertranslation that maps $\bar{u}$ cuts to $u=\bar{u}+f(x)$ cuts. $Y$ and $\bar{Y}$ are rotation BMS fields such that $Y$ is tangent to $u$ cuts and $\bar{Y}$ is tangent to $\bar{u}$ cuts.}
\centering
\includegraphics[scale=0.5]{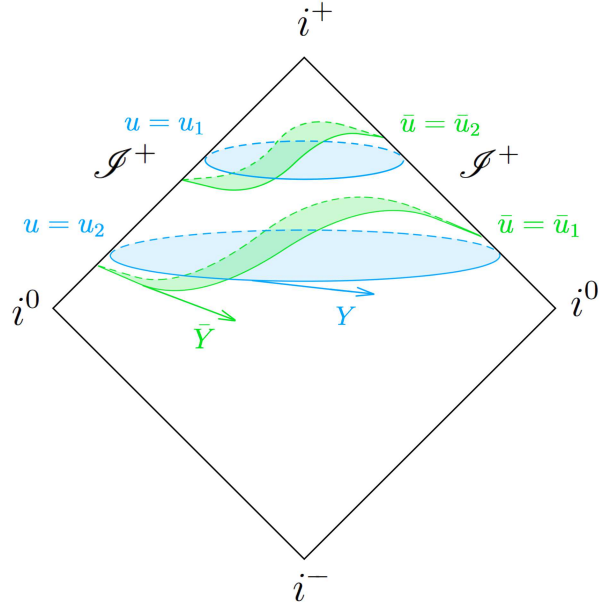}\end{figure}

We assume $\mathscr{I}^+$ extends from $i^0$ ($u=-\infty$) to $i^+$ ($u=+\infty$) and that there exists a constant $\varepsilon>0$ such that
\begin{equation}\label{news_decay}
N_{AB}( u,x) = O(|u|^{-1-\varepsilon}) \mbox{ as } u \rw \pm\infty.
\end{equation} The total flux of the angular momentum \eqref{AM} is defined to be \[\delta {J} (Y)=\lim_{u\rightarrow +\infty} {J}(u, Y)-\lim_{u\rightarrow -\infty} {J}(u, Y).\] 

When two Bondi-Sachs coordinates are related by a supertranslation, one shows that (see \eqref{supertranslation} below) \begin{equation}m(+)=\lim_{u\rightarrow + \infty} m(u, x)=\lim_{\bar{u} \rightarrow + \infty} \bar{m}(\bar{u}, x) \text{ and } m(-)=\lim_{u\rightarrow  -\infty} m(u, x)=\lim_{\bar{u} \rightarrow - \infty} \bar{m}(\bar{u}, x)\end{equation} are two functions on $S^2$.

\noindent \textit{Theorem 1- Under condition \eqref{news_decay}, suppose two Bondi-Sachs coordinate systems  are related by a supertranslation $f$, and $Y$ and $\bar{Y}$ are rotation BMS fields related by $f$ \eqref{BMS_related}. Then
\begin{equation}\label{flux_invariant}\delta {J}(\bar{Y})-\delta {J}(Y)= -\int_{S^2} \lt(2 f_{\ell\leq 1} \hat{Y}^A\nabla_A(m(+)-m(-)) \rt), \end{equation}
where $f=f_{\ell\leq 1}+f_{\ell\geq 2}$ is the decomposition into the corresponding harmonic modes.
}

\noindent \textit{Proof.}
The Einstein equation implies (see \cite{CJK, MW}):
\begin{equation}\label{Einstein} \begin{split}
\partial_u m&=-\frac{1}{8}N_{AB}N^{AB}+\frac{1}{4}\na^A\na^B N_{AB}\\
\partial_u N_A &= \na_A m -\frac{1}{4}\nabla^D(\nabla_D \nabla^E C_{EA}-\nabla_A \nabla^E C_{ED}) \\
&\quad +\frac{1}{4}\nabla_A(C_{BE} N^{BE})-\frac{1}{4}\nabla_B (C^{BD} N_{DA})+\frac{1}{2} C_{AB}\nabla_D N^{DB}.\end{split}\end{equation}

We calculate 
\begin{equation}\label{du_classical_AM}\begin{split}&\partial_u  \tilde{J}(u, Y)=\frac{1}{4}\int_{S^2}  Y^A  \lt[C_{AB}\nabla_{D}N^{BD} -N_{AB}\nabla_{D}C^{BD}-\nabla_B (C^{BD} N_{DA}) \rt]\end{split}\end{equation}

According to \eqref{AM}, the total flux $\delta J (Y)$ is thus 
\begin{equation}\begin{split}\label{AM_flux}\delta J(Y)=\delta \tilde{J}(Y) - \lt[\int_{S^2}Y^A  c\nabla_{A}m\rt]_{u=-\infty}^{u=+\infty},\end{split}  \end{equation}
where $ \delta \tilde{J}(Y)=\frac{1}{4}\int_{-\infty}^\infty \int_{S^2}Y^A  \lt[C_{AB}\nabla_{D}N^{BD} -N_{AB}\nabla_{D}C^{BD}-\nabla_B (C_{BD} N_{DA}) \rt] (u, \cdot) du$.

For a supertranslation \eqref{coord_change}, it is known that the mass aspect $\bar{m} (\bar u, x)
$, the shear $\bar{C}_{AB}(\bar u,x)$, and the news $ \bar{N}_{AB}(\bar u,x)$  in the $(\bar u, \bar{x})$ coordinate system are related to the mass aspect ${m} (u, x)
$, the shear ${C}_{AB}(u,x)$, and the news $ {N}_{AB}(u,x)$  in the $(u, x)$ coordinate system through:
\begin{equation}\label{supertranslation}\begin{split}
\bar{m}(\bar u,x) &= m(\bar u+f,x) + \frac{1}{2} (\na^B N_{BD})(\bar u+f,x) \na^D f \\
&\quad + \frac{1}{4} (\partial_{u}N_{BD})(\bar u+f,x) \na^B f\na^D f + \frac{1}{4} N_{BD}(\bar u+f,x) \na^B\na^D f\\
\bar{C}_{AB}(\bar u,x) &= C_{AB}(\bar u+ f(x),x) - 2 \na_A\na_B f + \Delta f \sigma_{AB} \\
\bar{N}_{AB}(\bar u,x) &= N_{AB}(\bar u+f(x),x)
\end{split}\end{equation}
See \cite[(C.117) and (C.119)]{CJK} for example. 

The decay condition of the news \eqref{news_decay} implies that the limits of the mass aspect and the shear satisfy
\begin{equation}\label{mass_shear_diff} \lim_{\bar u \rw \pm\infty} \bar {m}(\bar u,x) =  \lim_{u \rw \pm\infty}  m(u,x),
\lim_{\bar u\rw\pm\infty} \bar{C}_{AB}(\bar u,x) = \lim_{u\rw \pm\infty} C_{AB}(u,x) -2\na_A\na_B f + \Delta f \sigma_{AB}. 
\end{equation}
In particular, the limit of the potential $c$ satisfies:
\[\lim_{\bar u\rw\pm\infty} c(\bar u,x) = \lim_{u\rw \pm\infty} c(u,x) -2 f_{\ell\geq 2}. \]  It follows that the contribution from the second term of \eqref{AM_flux} in the difference $\delta {J} (\bar{Y})- \delta {J} (Y)$ is $\int_{S^2} [2f_{\ell\geq 2} \hat{Y}^A \na_A (m(+)-m(-))]$. On the other hand, the contribution from the first term in \eqref{AM_flux}, or $\delta \tilde{J} (\bar{Y})-\delta \tilde{J} (Y)$, after substituting \eqref{supertranslation}, integration by parts, and change of variables, is shown to be ( \cite{CKWWY}, see also \cite{ADK}) \[ \frac{1}{4} \int_{-\infty}^{+\infty} \lt[ \int_{S^2} f Y^A \na_A \big( N_{BD}N^{BD} - 2\na^B\na^D N_{BD} \big) \rt] du.\] At this point, we invoke the Einstein equation \eqref{Einstein} and rewrite the last integral as $\int_{S^2} [-2f \hat{Y}^A \na_A (m(+)-m(-))]$. Therefore $\delta J (\bar{Y})-\delta J (Y)$ is given by \eqref{flux_invariant}. By \eqref{EM}, the total flux of linear momentum is $\delta P^k=2\int_{S^2} (m(+)-m(-)) \tilde{X}^k$. It follows that 
\begin{equation}\label{AM_flux_transf}\begin{split}
\delta J (\bar{Y})  &=  \delta J (Y)+  \alpha_i \varepsilon^{ik}_{\;\;\;j} \delta P^j, \text{ if } f = \alpha_0 + \alpha_i \tilde X^i+f_{\ell\geq 2} \text{ and }\hat{Y}^A=\epsilon^{AB} \nabla_B\tilde{X}^k \end{split}
\end{equation} 
In particular, if $f$ is of harmonic mode $\ell\geq 2$, $\delta J (\bar{Y})=\delta J (Y)$ is invariant.

\section{Spacetime with vanishing news}

We consider a non-radiative spacetime in the sense that the news vanishes. This includes all model spacetimes such as Minkowski and Kerr. If in a Bondi-Sachs coordinate system $(u, x)$, $N_{AB}(u, x)$ vanishes, by \eqref{Einstein} the mass aspect $m(u,x)=\mathring{m}(x)$ is a function on $S^2$.

\noindent \textit{Theorem 2- Under the same assumption as Theorem 1, if in addition the news vanishes, then $J(\bar{u}, \bar{Y})\equiv J(\bar{Y})$ and $J(u, Y)\equiv J(Y)$ are independent of $\bar{u}$ and $u$, respectively, and are related by
\begin{equation}\label{constant_invariant}{J}(\bar{Y} )-J(Y)=-\int_{S^2} (2 f_{\ell\leq 1} \hat{Y}^A\na_A \mathring{m}),\end{equation}} where $\mathring{m}$ is the mass aspect.

\noindent \textit{Proof.} The vanishing of news also implies $C_{AB}$ and thus $c$  are independent of $u$.  The constancy of $J(u, Y)$ follows from \eqref{du_classical_AM}. We denote  $C_{AB}(u, x)=\mathring{C}_{AB}(x)$ and $c(u, x)=\mathring{c}(x)$. The exact formula for the angular aspect on a spacetime with vanishing news is obtained by integrating \eqref{Einstein} with respect to $u$:
\begin{equation}\label{AM_aspect1} N_A  (u, x)=  N_A  (u_0, x) +(u-u_0)(\na_A \mathring{m} -\frac{1}{4}\nabla^B \mathring{P}_{BA} )\end{equation} for any $u$ and fixed $u_0$,
where $\mathring{P}_{BA} =\nabla_B \nabla^E \mathring{C}_{EA}-\nabla_A \nabla^E \mathring{C}_{EB}$.

Suppose $(\bar{u}, \bar{x})$ is related to $(u, x)$ by a supertranslation $f$ \eqref{coord_change}.
By \eqref{supertranslation}, $\bar{N}_{AB}(\bar u,x)\equiv 0$ and ${J}(\bar{u}, \bar{Y})$ is independent of $\bar{u}$. In addition, their mass aspects, shears, and shear potentials are related by
\begin{equation}\label{transform}
\mathring{\bar{m}}= \mathring{m}, \mathring{\bar{C}}_{AB}=\mathring{C}_{AB} - F_{AB}, \text{ and } \mathring{\bar{c}}=\mathring{c}-2 f_{\ell\geq 2},
\end{equation} where $F_{AB}=2 \na_A\na_B f -\Delta f \sigma_{AB} $.

In this case, the angular aspect transforms according to \cite[(C.123)]{CJK}  
$\bar{N}_A(\bar{u}, x)=N_{A}(\bar{u}+f, x)+3\mathring{m}\nabla_A f-\frac{3}{4}\mathring{P}_{BA} \na^B f$ 
\footnote{Note that the convention of angular momentum aspect in \cite{CJK} is $-3N_A$.}. Combining this with \eqref{AM_aspect1} and set $u=\bar{u}+f$, we obtain 
\begin{equation}\label{AM_aspect2} \bar{N}_A  (\bar{u}, x)=  N_A  (u_0, x) +(\bar{u}-u_0+f)       (\na_A \mathring{m} -\frac{1}{4}\nabla^B \mathring{P}_{BA} )+3\mathring{m}\nabla_A f-\frac{3}{4}\mathring{P}_{BA} \na^B f\end{equation} for any $\bar{u}$ and fixed $u_0$. 

Let ${J}(\bar{u}_0, \bar{Y})$ (resp. $J(u_0, Y)$) be the angular momentum of the $\bar{u}=\bar{u}_0$ (resp. $u=u_0$) cut in the $(\bar{u}, \bar{x})$ (resp. $(u, x)$) coordinate system. Taking their difference and applying \eqref{transform}, 
\[\begin{split}{J} (\bar{u}_0, \bar{Y})-J(u_0, Y)&=\int_{S^2} Y^A \lt[\bar{N}_A  (\bar{u}_0, x)-  N_A  (u_0, x)\rt]\\
&\quad +\frac{1}{4}\int_{S^2}  Y^A \lt[  \mathring{C}_{AB}\nabla_D F^{BD}+F_{AB} \nabla_D \mathring{C}^{BD}- F_{AB} \nabla_D {F}^{BD}    \rt]\\
&\quad + \int_{S^2} (2f_{\ell\geq 2} \hat Y^A\na_A \mathring{m}).\end{split}\]
We  then apply \eqref{AM_aspect2} to show that the sum of the first two integrals on the above right hand side is  $\int_{S^2} (-2 f \hat{Y}^A \na_A \mathring{m} )$ and ${J}(\bar{Y})-J(Y)$ is of the desired expression \eqref{constant_invariant}, see \cite{CKWWY} for details.  In particular,\begin{equation}\label{AM_transf}
 J (\bar{Y}) = J (Y)+  \alpha_i \varepsilon^{ik}_{\;\;\;j} P^j, \text{ if } f = \alpha_0 + \alpha_i \tilde X^i+f_{\ell\geq 2} \text{ and }\hat{Y}^A=\epsilon^{AB} \nabla_B\tilde{X}^k. 
\end{equation}

\section{Conserved quantities on null infinity}

The limit of the quasilocal center of mass defined in \cite{CWY2} also gives a new definition of center of mass at null infinity \cite{KWY-CWY1}.  We say that a BMS field $Y$ is a {\it boost BMS field} if in a Bondi-Sachs coordinate system $(u, x)$,
\begin{equation}\label{boost_BMS} Y=\hat{Y}^A\frac{\partial}{\partial x^A}+u\tilde{X} \frac{\partial}{\partial u}, \end{equation} where $\hat{Y}^A=\nabla^A \tilde{X}$ and $\tilde{X} $ is a function on $S^2$ of harmonic mode $\ell=1$. 

\noindent \textit{Definition of center of mass:} For a boost BMS field $Y$ \eqref{boost_BMS}, the center of mass of a $u$ cut is defined to be:
\begin{align}\label{COM}
\begin{split}
 C(u, Y)&= \int_{S^2} \nabla^{A}\tilde{X} \lt(N_A - \frac{1}{4} C_{AB} \na_D C^{DB} - \frac{1}{16} \na_A \lt( C^{DE}C_{DE}\rt)\rt) (u, \cdot)-2u\int_{S^2}(\tilde{X} m) (u, \cdot) \\
&+\int_{S^2} c\lt(3 \tilde X m - \na^A \tilde{X}  \na_A m\rt) (u, \cdot)\\
& +\int_{S^2} \lt( 2 \nabla^{A}\tilde{X} \epsilon_{AB} (\na^B \cb) m      - \frac{1}{16} \tilde X \na_A (\Delta+2)\cb \na^A (\Delta+2)\cb\rt) (u, \cdot)\\
\end{split}
\end{align}

Suppose $(\bar{u}, \bar{x})$ is related to $(u, x)$ by a supertranslation $f$ \eqref{coord_change}. Let $\bar{Y}$ be the boost BMS field
\[\bar{Y}=\hat{Y}^A\frac{\partial}{\partial \bar{x}^A}+\bar{u}\tilde{X} \frac{\partial}{\partial \bar{u}},\] for the same $\hat{Y}^A$ in \eqref{boost_BMS}. $\bar{Y}$ and $Y$ \eqref{boost_BMS} are said to be related by the supertranslation $f$ and
${C}(\bar{u}, \bar{Y})$  is obtained by replacing $m, C_{AB}, N_{AB}, N_A, c, \underline{c}, u$ in \eqref{COM} with $\bar{m}, \bar{C}_{AB}, \bar{N}_{AB}, \bar{N}_A, \bar{c}, \bar{\underline{c}}, \bar{u}$. We show in \cite{CKWWY} that their total fluxes are related by 
\[\delta {C} (\bar{Y})-\delta C (Y)= -2 \int_{S^2} f_{\ell\le1} \na^A \tilde X \na_A \big( m(+) - m(-) \big) + 6 \int_{S^2} f_{\ell\le1} \tilde X \big( m(+) - m(-) \big),\] and therefore,
\begin{equation} \label{COM_flux_transf}
\delta {C}(\bar{Y}) - \delta C(Y)
= \alpha_0 \delta P^k  + \alpha_k \delta E  \text{ if } f = \alpha_0 + \alpha_i \tilde X^i+f_{\ell\geq 2} \text{ and }\hat{Y}^A= \nabla^A\tilde{X}^k .\end{equation}

When the spacetime has vanishing news, we show in \cite{CKWWY} that
${C}(\bar{u}, \bar{Y})\equiv {C} (\bar{Y})$ and $C(u, Y)\equiv C (Y)$ are independent of $\bar{u}$ and $u$, respectively, and are related by
\begin{equation}\label{COM_transf}{C}(\bar{Y})-C(Y)=-2 \int_{S^2} f_{\ell\le1} \na^A \tilde X \na_A \mathring{m} + 6 \int_{S^2} f_{\ell\le1} \tilde X \mathring{m}.       \end{equation} 

Fixing $\tilde{X}^k, k=1, 2, 3$ and denoting $J^k(u)\coloneqq J(u, Y)$ for $Y=\epsilon^{AB}\nabla_B \tilde{X}^k \frac{\partial}{\partial x^A}$ and $C^k(u)\coloneqq C(u, Y)$ for $Y=\nabla^A \tilde{X}^k\frac{\partial}{\partial x^A}+u\tilde{X}^k\frac{\partial}{\partial u}  $, we obtain the new definitions of angular momentum $J^k(u)$ and center of mass $C^k(u)$. They complement the classical Bondi-Sachs energy momentum $E(u), P^k(u)$ and form a set of conserved quantities that correspond to the Poincar\'e symmetry. All of them can be derived from the limits of quasilocal conserved quantities defined in \cite{WY, CWY2}. The Poincar\'e symmetry is due to the choice of Minkowski reference and is acquired through the reference embedding into the Minkowski spacetime \cite{WY}.
The use of the Minkowski reference is essential. In the early days of the study of angular momentum flux, there was confusion about nonzero flux in the Minkowski spacetime, which was eventually clarified in \cite{AS}. For our definitions of angular momentum and center of mass, not only that the fluxes are zero, but also that the angular momentum and center of mass are zero in any Bondi-Sachs coordinate system of the Minkowski spacetime.

\section{Conclusions}

We obtain a complete set of ten conserved quantities $(E, P^k, J^k, C^k)$ at null infinity (all as functions of the retarded time $u$) that satisfy the following properties:

(1) $(E, P^k, J^k, C^k)$ all vanish for any Bondi-Sachs coordinate system of the Minkowski spacetime. 

(2) In a Bondi-Sachs coordinate system of the Kerr spacetime, $P^k$ and $C^k$ vanish, and $E$ and $J^k$ recover the mass and angular momentum. $(E, P^k, J^k, C^k)$ are supertranslation invariant. 

(3) If a spacetime admits a Bondi-Sachs coordinate system such that the news vanishes, then $(E, P^k, J^k, C^k)$ are constant (independent of the retarded time $u$) and supertranslation  invariant. 

(4) On a general spacetime,  the total fluxes of $(E, P^k, J^k, C^k)$ are supertranslation invariant. 

(5) $(E, P^k, J^k, C^k)$ and their fluxes transform according to basic physical laws \eqref{AM_flux_transf},
 \eqref{COM_flux_transf}, \eqref{AM_transf}, \eqref{COM_transf} under ordinary translations. 


Under additional assumptions and $\nabla_A m(+)=\nabla_A m(-)=0$, the supertranslation invariance holds by Remark 2 of  the second paper in \cite{ADK}. This manifests the importance of the correction term (the last term) in our definition of angular momentum \eqref{AM} which is supertranslation invariant without any additional assumptions. 
We only consider supertranslations which correspond to fixing the 2-metric $\sigma_{AB}$ at $\mathscr{I}^+$. The transformation of angular momentum and fluxes under boosts, which change $\sigma_{AB}$ by conformal factors,  will be discussed in a forthcoming
work.

\begin{acknowledgements}P.-N. Chen is supported by Simons Foundation collaboration grant \#584785, M.-T. Wang is supported by NSF grant DMS-1810856, Y.-K. Wang is supported by Taiwan MOST grant 109-2628-M-006-001-MY3 and S.-T. Yau is supported by. The authors would like 
to thank Professor Abhay Ashtekar for helpful discussions. \end{acknowledgements}


\end{document}